\documentclass[12pt]{iopart}
\usepackage{amsfonts}
\usepackage{amssymb}
\usepackage{epsfig}
\usepackage{subfigure}
\usepackage{graphicx}

\def\beq{\begin{equation}}
\def\eeq{\end{equation}}
\def\beqa{\begin{eqnarray}}
\def\eeqa{\end{eqnarray}}
\def\hf{\textstyle{1\over2}}
\def\3hf{\textstyle{\frac{3}{2}}}

\newcommand{\ket}[1]{\vert #1 \rangle}

\newcommand{\bra}[1]{\langle #1 \vert}
\newcommand\unit{\mathinner{\hbox{1}\mkern-4mu\hbox{l}}}

\newcommand\ip[2]{\langle #1\,\vert\,#2\rangle}

\renewcommand{\e}{\hbox{\rm e}}
\newcommand{\myfrac}[2]{\leavevmode\kern.1em\raise.5ex\hbox{\scriptsize
$#1$}\kern-.1em {\scriptsize
/}\kern-0.10em\lower.25ex\hbox{\scriptsize $#2$}}

\begin{document}

\title[$SU(3)$ Phase operators from FFT]
{$SU(3)$ phase states and finite Fourier transform}
\author{Brandon Zanette and Hubert de Guise}
\address{Department of Physics, Lakehead University,
Thunder Bay, Canada}

\begin{abstract}
We describe the construction of $SU(3)$ phase operators using Fourier-like transform on a hexagonal lattice.  The advantages and disadvantages of this approach are contrasted with other results, in particular with the more traditional approach based on polar decomposition of operators.
\end{abstract}

\section{Introduction: complementarity and the Fourier transform}

The idea of complementarity in quantum mechanics goes back to Bohr and his attempt
to explain wave-particle duality.  The concept was sharpened by
Pascual Jordan, who stated \cite{Jammer} that
\begin{center}
\emph{For a given value of x all values of p are equally possible.}
\end{center}
This formulation automatically singles out the Fourier transform connecting operators like $\hat x$ and $\hat p$ as their respective (generalized) eigenstates satisfy
\beq
\ip{x}{p}\sim \e^{ixp/\hbar} \quad \Rightarrow \quad \vert \ip{x}{p}\vert^2 =\hbox{\rm constant}
\eeq
The concept is not limited to continuous systems but also exists in finite dimensions.  In this contribution, we will discuss the construction of $SU(3)$ phase operators, which are expected to be complementary to number operators.  This contribution emphasizes the importance of the finite Fourier transform, and in particular a new type of generalization of the Fourier transform that is constructed to preserve the symmetry of a hexagonal lattice, which is the natural (discrete) lattice to describe states appropriate to the description of a collection of three--level systems.  Our approach should be contrasted with the approach of Dirac \cite{Dirac}, which emphasizes polar decompositions, and which has been applied to
$SU(2)$ and other systems in \cite{su2phasepapers} and \cite{Kibler}.

\section{Two examples}

Consider first a spin-$\hf$ system, taking as operators the Pauli matrices
$\sigma_x,\sigma_y$ and $\sigma_z$.   The eigenstates $\{\ket{+}_z,\ket{-}_z\}$ of $\sigma_z$ and the eigenstates
$\{\ket{+}_x,\ket{-}_x\}$ of $\sigma_x$  are complementary:
\beq
\vert _x\ip{+}{+}_z\vert^2=\vert _x \ip{-}{+}_z\vert^2
=\vert _x \ip{+}{-}_z\vert^2=\vert _x \ip{-}{-}_z\vert^2 = \hf=\hbox{\rm constant}\, .
\eeq
The eigenstates of $\sigma_z$ and $\sigma_x$ are related by a \emph{finite} Fourier transform:
\beq
F=\frac{1}{\sqrt{2}}\left(
{\renewcommand{\arraystretch}{0.75}
\renewcommand\arraycolsep{0.25em}\begin{array}{cc} 1& 1 \\ 1 & -1\end{array}}\right)
\eeq
The operators $\sigma_x$ and $\sigma_z$ are said to be complementary.
The same property holds for the pair $\sigma_y$ and $\sigma_z$ and for the pair $\sigma_y$ and $\sigma_x$.
The transformation matrix connecting any two sets of eigenstates remains a finite Fourier transform.

A similar construction exists for a three--level system (or qutrit).  Defining 
\beq
\hat Z=\left(
{\renewcommand{\arraystretch}{0.75}
\renewcommand\arraycolsep{0.25em}\begin{array}{ccc} 1 &0&0\\ 0&\omega&0 \\ 0& 0& \omega^2\end{array}}\right)\, ,\qquad
\hat X=\left(
{\renewcommand{\arraystretch}{0.75}
\renewcommand\arraycolsep{0.25em}\begin{array}{ccc} 0 &1&0\\0& 0&\omega \\ \omega^2 & 0 &0 \end{array}}\right)\, ,
\qquad\omega=\e^{2\pi i /3}\, , \label{Xqutrit}
\eeq
and writing their respective eigenstates as $\{\ket{0_z},\ket{1_z},\ket{2_z}\}$ and 
$\{\ket{0_x},\ket{1_x},\ket{2_x}\}$  we find for instance
$
\vert \ip{1_x}{0_z}\vert^2 =\vert \ip{2_x}{2_z}\vert^2=\textstyle\frac{1}{3}
$
with all other such overlaps constant.  Here again, the eigenstates of $\hat X$ and $\hat Z$ are related by a finite Fourier transform:
\beq
F=\frac{1}{\sqrt{3}}
\left(
{\renewcommand{\arraystretch}{0.75}
\renewcommand\arraycolsep{0.25em}\begin{array}{ccc}
\omega & 1  & \omega^2 \\
\omega & \omega^2 & 1 \\
1 & 1 & 1 \end{array}}\right)\, ,\label{FSU3}
\eeq

This is a good point to mention some of the properties of the finite Fourier matrix $F$.  It is unitary, which implies
    \beq
    \sum_i F_{ki}^* F_{ij}=\delta_{ij}
    \eeq
(This would be orthogonality under integration in the continuous case.)
$F^4=\unit$, and its entries are characters of finite Abelian groups.  In dimension $n$: 
    \beq
    F_{jk}= {\e^{2\pi i j k/n}}/{\sqrt{n}} \, .
    \eeq
    Finally, $\vert F_{ij}\vert$ have constant magnitude, connecting with Jordan's definition of complementarity.

\section{$SU(2)$ phase states}

Following Dirac \cite{Dirac} and others \cite{su2phasepapers}, phase operators in $su(2)$ (and other) systems
are constructed by writing the matrix for $\hat S_+$ (or $\hat S_-$) in polar form, {\it viz.}
\beq
\hat S_+\mapsto \displaystyle \sum_{m=0}^{j-1} c_m\ket{j,m+1}\bra{j,m}=E\cdot D
\eeq
where $D$ is  diagonal and $E$
is a ``phase'' part, containing entries which produce the shifting action of $S_+$ on the basis states. The operator $E$ is expected on physical grounds to be complementary to 
the diagonal operator $\hat S_z$.

Geometrically, the set of eigenvalues $\{m\}$ of $\hat S_z$ acting on number states $\ket{jm}$ are equidistant  points on a line
and the action of the ladder operators $\hat S_{\pm}$ takes a point $m$ to its neighbor $m\pm 1$. The action of $S_+$ is pictorially represented in Fig.\label{s+shift}.
\begin{figure}[h!]
\begin{center}
\includegraphics[scale=0.45]{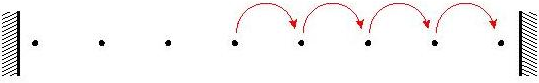}
\end{center}
\caption{The shift action of the $su(2)$ raising operator $\hat S_+$.}\label{s+shift}
\end{figure}

Because  $\ket{jj}$ is killed by $\hat S_+$, the rank of $\hat S_+$  is one less than the dimension of the system,
so that  $E$ is not completely defined: we can adjust the entries in one line.  The usual choice makes $E$ cyclic ($ E^{2j+1}=\unit$) so it generates an Abelian group of order $2j+1$: 
\beq
E=\e^{i\varphi}\mapsto\left(
\begingroup
\everymath{\scriptstyle}
\small
{\renewcommand{\arraystretch}{0.65}
\renewcommand\arraycolsep{0.15em}
\begin{array}{ccccc}
0&1&\ldots& & \\
 &0&1     & & \\
 & &\ddots& & \\
 & &      &0&1\\
1&0&\ldots&0&0\end{array}
}
\endgroup
\right)
=\sum_{m=-j}^{j-1}\ket{m+1}\bra{m}+ \ket{-j}\bra{j}\, .
\eeq
This $E$ is unitary, and can be written in the form $E=\e^{i\hat \varphi}$, with
$\hat\varphi$ the putative hermitian phase operator.  The eigenvectors of $E$ are eigenvectors of $\hat\varphi$ and {\it defined} to be the $SU(2)$ phase states.
The components of the $m$'th eigenvector $\ket{\varphi_m}$ are just elements of a Fourier matrix $F$.
Thus, the phase eigenstate $\ket{\varphi_m }$ is given by
\beq
\ket{\varphi_m}=\sum_k F_{mk}\ket{jk}
= \textstyle\frac{1}{\sqrt{2j+1}}\displaystyle\sum_k\,\e^{2\pi i k m/(2j+1)} \ket{jk}\, .
\eeq

\section{$SU(3)$ and $SU(3)$ phase states}

\subsection{Geometry of $SU(3)$ states}

\begin{figure}[h!]
\begin{center}
\includegraphics[scale=2.2]{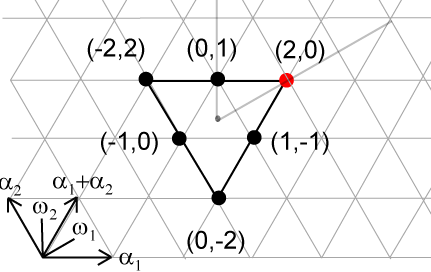} \qquad\  \includegraphics[scale=0.29]{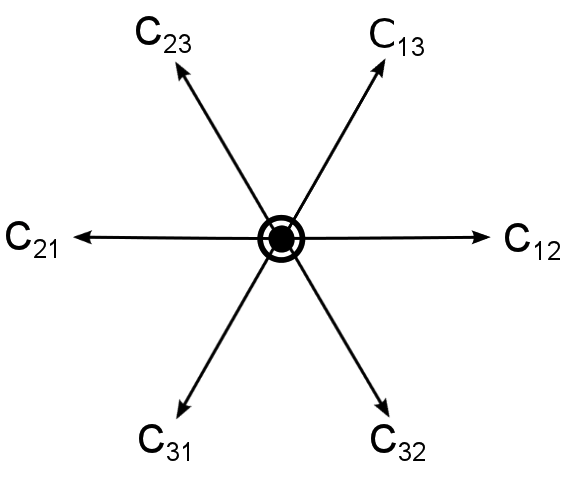}
\caption{Left: The population differences $(n_1-n_2,n_2-n_3)$ for the
states $\ket{n_1n_2n_3}$ with $n_1+n_2+n_3=2$.  $(a,b)$ is located at 
$a\omega_1+b\omega_2$ with $\omega_1, \omega_2$ the lattices vectors.  Right:
The graphical representation of the shift action of the ladder generators
of $su(3)$ on the hexagonal lattice.  The ringed dot at the center represents the two diagonal population difference operators $\hat h_1$ and $\hat h_2$, which do not shift the basis
states.}\label{20weights}
\end{center}
\end{figure}

The algebra $su(3)$  appears naturally in the construction of number--preserving transition operators for three--level systems. 
There are six transition operators, usually denoted by $\hat C_{ij}=a_i^\dagger a_j$ for
$i\ne j=1,2,3$, and two population differences
$\hat h_1=a_1^\dagger a_1 - a_2^\dagger a_2$ and $\hat h_2=a_2^\dagger a_2 -a_3^\dagger a_3$.
The states $\ket{200}$ and $\ket{110}$, for instance, correspond to
the pairs $(2,0)$ and $(0,1)$ of population differences.   Pairs are located on a hexagonal lattice with basis vectors $\omega_1$ and $\omega_2$
as illustrated on the left of  Fig.\ref{20weights}.

The action of $\hat C_{ij}$ on lattice points is illustrated on the right of Fig. \ref{20weights}.  Basis vectors $\alpha_1$ and $\alpha_2$ associated to the operators 
$\hat C_{12}$ and $\hat C_{23}$ respectively are dual (reciprocal) to the lattice vectors $\omega_1$ and $\omega_2$, respectively, as illustrated.
Using the hexagonal geometry, two points $(a,b)$ and $(c,d)$ corresponding to two pairs of population differences differ by an integer combination of the vectors $\alpha_1$ and $\alpha_2$.  The action of 
$\hat C_{ij}$ on the state $\ket{n_1n_2n_3}$ is to translate the point
$(n_1-n_2,n_2-n_3)$ by the vector associated to $\hat C_{ij}$ to the point $(n'_1-n'_2,n'_2-n'_3)$, so that, for instance
\beq
\hat C_{12}\ket{110}\sim \ket{200}\, \Rightarrow
(0,1)\mapsto  (2,0)=\alpha_1+(0,1)\, .
\eeq
The central ringed dot represents the two diagonal population difference operators
$\hat h_1$ and $\hat h_2$.
There should be one phase operator conjugate to each $\hat h_i$.

\subsection{Two solutions; boundaries}

If we approach the construction of $SU(3)$ phase states using polar decompositions, we are faced with an interesting problem.  Because there are two basic
shift directions, $\alpha_1$ and $\alpha_2$, each one of $\hat C_{12}$ and $\hat C_{23}$ will come with its own set of not necessarily mutually compatible boundary conditions.

In the simplest case of the states $\ket{100},\ket{010}$ and $\ket{001}$,  the shift matrices $E_{12}$ and $E_{23}$ that enter in the decompositions of $\hat C_{12}$ and
$\hat C_{23}$ respectively contain two lines that cannot be uniquely determined.  These matrices can be completed in two different ways.
First, we can write
\beqa
E_{12}&=&\left(
{\renewcommand{\arraystretch}{0.65}
\renewcommand\arraycolsep{0.15em}\begin{array}{ccc} 0&1&0\\ 1&0&0 \\ 0&0&1\end{array}}\right) 
=\left[\ket{010}\bra{100}+\ket{100}\bra{010}\right]+\ket{001}\bra{001}\, , \label{E12su2}\\
E_{23}&=&\left(
{\renewcommand{\arraystretch}{0.65}
\renewcommand\arraycolsep{0.15em}
\begin{array}{ccc} 1&0&0\\ 0&0&1 \\ 0&1&0\end{array}}\right)=\ket{100}\bra{100}+\left[\ket{010}\bra{001}+\ket{001}\bra{010}\right]\, . \label{E23su2}
\eeqa
\begin{figure}[h!]
\begin{center}
\includegraphics [scale=0.4]{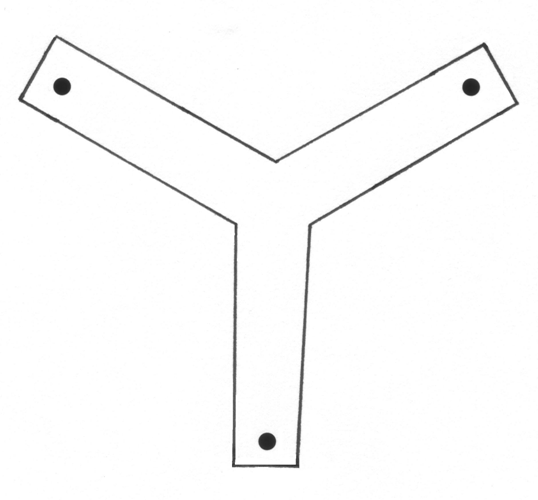}\qquad\qquad
\includegraphics [scale=0.32]{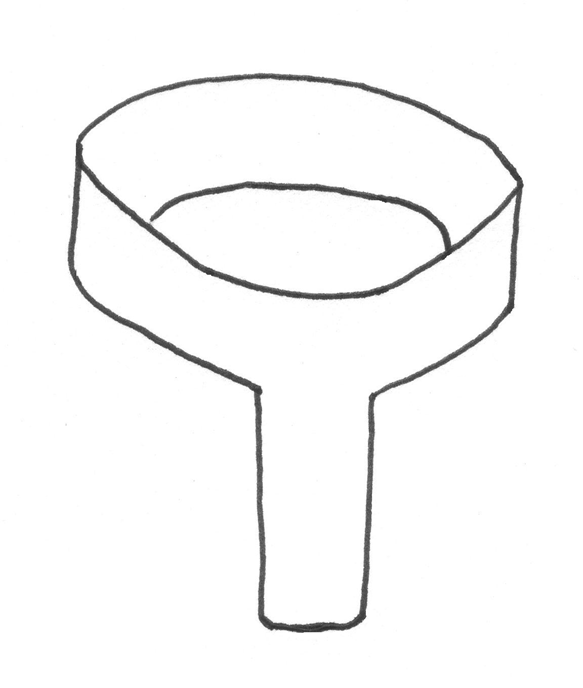}
\caption{A graphical representation of how cyclic boundary conditions can be imposed to complete the matrix $E_{12}$ to yield Eqn.(\ref{E12su2}). }
\end{center}
\end{figure}

This kind of solution also exists for more general cases where $n_1+n_2+n_3>1$.  It consists in considering subsets of states with the same value
for $n_3$ - such subsets of states fall on lines parallel to the $\alpha_1$ direction -  
and following the procedure of $SU(2)$ on each lineto obtain $E_{12}$.  Similarly, by considering subsets of 
states with the same value of $n_1$ - these states now lie on lines parallel to the $\alpha_2$ direction - we can follow the $SU(2)$ procedure for each line
and obtain $E_{23}$.  However, one feature of this solution, already present in Eqns.(\ref{E12su2}) and (\ref{E23su2}), is that the unitary phase operators do not
commute:
\beq
[E_{12},E_{23}]\ne 0 \quad\Rightarrow \e^{i\theta \hat\varphi_{12}}\e^{i \gamma\hat \varphi_{23}}\ne \e^{i(\theta \hat\varphi_{12}+ \gamma\hat \varphi_{23})}
\eeq
This in turns implies that the phases are not additive.

For the case of the states $\ket{100},\ket{010}$ and $\ket{001}$, it is possible to find shift matrices $E_{12}$ and $E_{23}$ compatible with the polar decomposition of
the respective operators and so that $[E_{12},E_{23}]=0$.  These matrices are
\beq
E_{12}=\left(
{\renewcommand{\arraystretch}{0.65}
\renewcommand\arraycolsep{0.15em}\begin{array}{ccc} 0&1&0\\ 0&0&\omega \\ \omega^2 &0&0\end{array}}\right) 
\, , \qquad 
E_{23}=\left(
{\renewcommand{\arraystretch}{0.65}
\renewcommand\arraycolsep{0.15em}
\begin{array}{ccc} 0&\omega^2&0\\ 0&0&1 \\ \omega&0&0\end{array}}\right)\, ,\qquad \omega=\e^{2\pi i /3}\, . \label{E12wrap}
\eeq
\begin{figure}[h!]
\begin{center}
\includegraphics [scale=0.4]{weight10.png}\qquad\qquad
\includegraphics [scale=0.375]{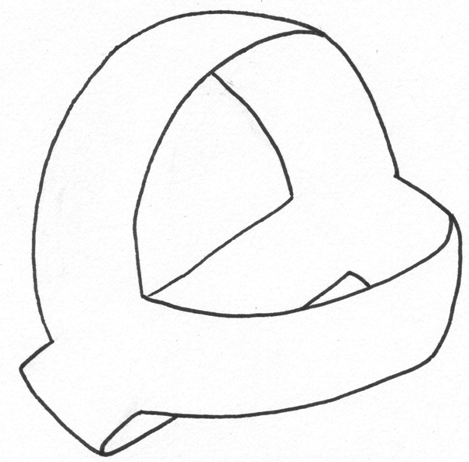}
\caption{A graphical representation of how cyclic boundary conditions can be imposed to complete the matrices $E_{12}$ and $E_{23}$ to yield Eqn.(\ref{E12wrap}). }
\end{center}
\end{figure}
Note that $E_{12}$ is just the operator $\hat X$ of Eqn.(\ref{Xqutrit}) while $E_{23}=\hat X^2$.  Clearly, $E_{12}$ and $E_{23}$ commute.  However, we have not been able to find
similar solutions for sets of states with $n_1+n_2+n_3>1$.

\subsection{Finite Fourier transform on a hexagonal lattice}

As an alternative to the construction based on polar decomposition, we look for a finite Fourier transform adapted to the discrete hexagonal symmetry natural to $SU(3)$ states.  Such an FFT was proposed in \cite{Patera}  and will be adapted to our needs.

\begin{figure}[h!]
\begin{center}
\includegraphics[scale=1.5]{Weight20}
\qquad
\includegraphics[scale=1.75]{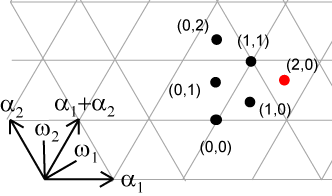}
\caption{An example of how the pairs of populations differences $(p,q)$ obtained from the states $\ket{n_1n_2n_3}$ with
$n_1+n_2+n_3$ are mapped to the first hextant.}\label{mappedset}
\end{center}
\end{figure}

\begin{figure}[h!]
\begin{center}
\includegraphics[scale=1.5]{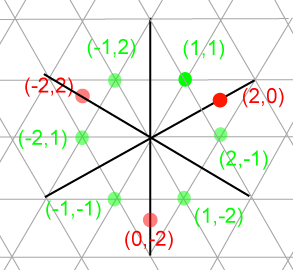}
\end{center}
\caption{The orbits of the points $(2,0)$ (in red) and $(1,1)$ (in green).}\label{orbits}
\end{figure}

We start with the physical states $\ket{n_1n_2n_3}$.  The procedure of \cite{Patera}  requires that the ``data points'' be in the first hextant of the lattice, so we find a rigid displacement of the set of population differences $(n_1-n_2,n_2-n_3)$ corresponding to the physical states so that every pair $(n_1-n_2,n_2-n_3)$ is mapped to a single point in the first hextant.  One can show that the rigid displacement is a linear transformation comprising a translation, a rotation and a change of scale of the original pairs of points.  The final result of the sequence is
\beq
\ket{n_1n_2n_3}\mapsto (n_1-n_2,n_2-n_3)\mapsto (n_1,n_2)
\eeq
An example of result is given in Fig.\ref{mappedset}.
We obtain for each point $(a,b)$ in the first extant its orbit, i.e. the set of points obtained by considering reflections of $(a,b)$ through mirrors 
perpendicular to $\alpha_1$ and $\alpha_2$.  Depending on the value of 
$a$
and $b$, an orbit may contain $1$, $3$ or $6$ points.   The orbit for the points $(2,0)$ and $(1,1)$ are illustrated in Fig.\ref{orbits}.
Each orbit is labelled by its starting point $(a,b)$ in the first extant.  There is the same number of orbits as the number of states.  
Each orbit is used to construct a so--called orbit function
\beqa
&&\chi_{(a,b)}(n_1,n_2)\sim
\omega^{(2a+b)n_1+(a+2b)n_2}+\omega^{(b-a)n_1+(a+2b)n_2}+\omega^{(2a+b)n_1+(a-b)n_2}\nonumber \\
&&\qquad\  +\omega^{-(a-b)n_1-(b+2a)n_2} +\omega^{-(a+2b)n_1+(b-a)n_2}+\omega^{-(2b+a)n_1-(b+2a)n_2}\, ,
\eeqa
with $(n_1,n_2)$ points in the first hextant.  The functions $\chi$ as closely related to characters of elements of finite order of $SU(3)$.

It is \emph{essential} to rigidly translate  the population differences of physical states.  Two states
$\ket{n_1n_2n_3}$, $\ket{n_1'n_2'n_3'}$ which differ only by a permutation of $n_1,n_2,n_3$ yield population differences 
$(n_1-n_2,n_2-n_3)$ and $(n_1'-n_2',n_2'-n_3')$ that are 
\emph{on the same orbit} and so produce identical functions $\chi$.  It is only once the population differences have been translated to the first hextant  that 
$(n_1,n_2)$ and $(n_1',n_2')$ will lie on different orbits.

The functions $\chi$ need to be properly normalized and weighted as described in \cite{Patera}  but, once this is done, they satisfy an orthogonality relation
\beq
\sum_{n_1,n_2}\left(\chi_{(a,b)}(n_1,n_2)\right)^*\,\chi_{(a',b')}(n_1,n_2)\sim\delta_{aa'}\delta_{bb'}\, .
\eeq
The orbit functions can then be used to obtain a Fourier matrix
\beq
    F=\left(
    {\renewcommand{\arraystretch}{0.65}
\renewcommand\arraycolsep{0.15em}\begin{array}{ccc}
    \chi_{a_1,b_1}(s_1,s_2)_1&\ldots&\chi_{a_1,b_1}(s_1,s_2)_k\\
    \vdots & \ddots & \vdots \\
    \chi_{a_k,b_k}(s_1,s_2)_1&\ldots&\chi_{a_k,b_k}(s_1,s_2)_k
    \end{array}}\right)
\eeq
So defined the matrix $F$ immediately satisfies the majority of the conditions given at the end of Sec.2.  In particular, for the set of states $\{\ket{100},\ket{010},\ket{001}$, the matrix $F$
is exactly that of Eqn.(\ref{FSU3}).   However, for other states with $n_1+n_2+n_3>1$, the matrix $F$ no longer contains entries of the same magnitude.  For instance, using the
states $\ket{n_1n_2n_3}$ with $n_1+n_2+n_3=2$, we find
\beq
F=\frac{1}{2}\left(
{\renewcommand{\arraystretch}{1.1}
\renewcommand{\arraycolsep}{0.195em}
\begin{array}{cccccc}
\frac{1}{\sqrt{3}} & 1 & \frac{1}{\sqrt{3}} & 1 & 1 & \frac{1}{\sqrt{3}} \\
1 & -\frac{1}{\sqrt{3}}\,\omega&\omega^2 & -\frac{1}{\sqrt{3}}\,\omega^2&\frac{-1}{\sqrt{3}}&\omega \\
\frac{1}{\sqrt{3}} & \omega^2 & \frac{1}{\sqrt{3}}\,\omega
& \omega & 1&\frac{1}{\sqrt{3}}\,\omega \\
1 & -\frac{1}{\sqrt{3}} & \omega & -\frac{1}{\sqrt{3}}\,\omega& \frac{-1}{\sqrt{3}} &\omega^2 \\
1 & \frac{-1}{\sqrt{3}} & 1 & \frac{-1}{\sqrt{3}} & \frac{-1}{\sqrt{3}} & 1\\
\frac{1}{\sqrt{3}} & \omega & \frac{1}{\sqrt{3}}\,\omega^2
& \omega^2 & 1 & \frac{1}{\sqrt{3}}\,\omega
\end{array}}
\right)\, ,\qquad\omega=\e^{2\pi i/3}\, .
\eeq

\subsection{$SU(3)$ phase states}

Now \emph{define} $SU(3)$ phase states as transforms of the shifted population difference eigenstates:
\beq
    \ket{\eta_1,\eta_2}_{(n_1,n_2)}\equiv \sum_{t_1,t_2}F_{(n_1,n_2),(t_1,t_2)}\ket{t_1,t_2}
\eeq
Phase ``operators'' are conjugate to population difference operators:
\beq
\hat\eta_1=F\hat h_1\,F^{-1}\, ,\qquad \hat \eta_2=F\hat h_2 F^{-1} \, . \label{su3phaseops}
\eeq
Since  $[\hat h_1,\hat h_2]=0$, we recover $[\hat\eta_1,\hat \eta_2]=0$: phases commute.

\subsection{Complementarity and number difference distributions}

To get insight into what phase states ``look like''  we consider the probabilities $\vert \ip{n_1,n_2}{\eta_1,\eta_2}\vert^2$.
Recall  the correspondences $\ket{n_1n_2n_3}\leftrightarrow (n_1,n_2)$ between physical states and their translated population difference.  Thus
\beq
\ket{100}\leftrightarrow (1,0)\qquad
\ket{010}\leftrightarrow (0,1)\qquad
\ket{001}\leftrightarrow (0,0)\, .
\eeq
For these states we find, for every $(n_1,n_2)$ and every $(a,b)$:
\beq
\vert\bra{(n_1,n_2)}\left[F\ket{(a,b)}\right]\vert^2=\textstyle\frac{1}{3}\nonumber \\
\eeq
This is no surprise as $\vert F_{(n_1n_2),(a,b)}\vert^2 = \frac{1}{3}$ for this case.

\begin{figure}[h!]
\begin{center}
\includegraphics[scale=0.4]{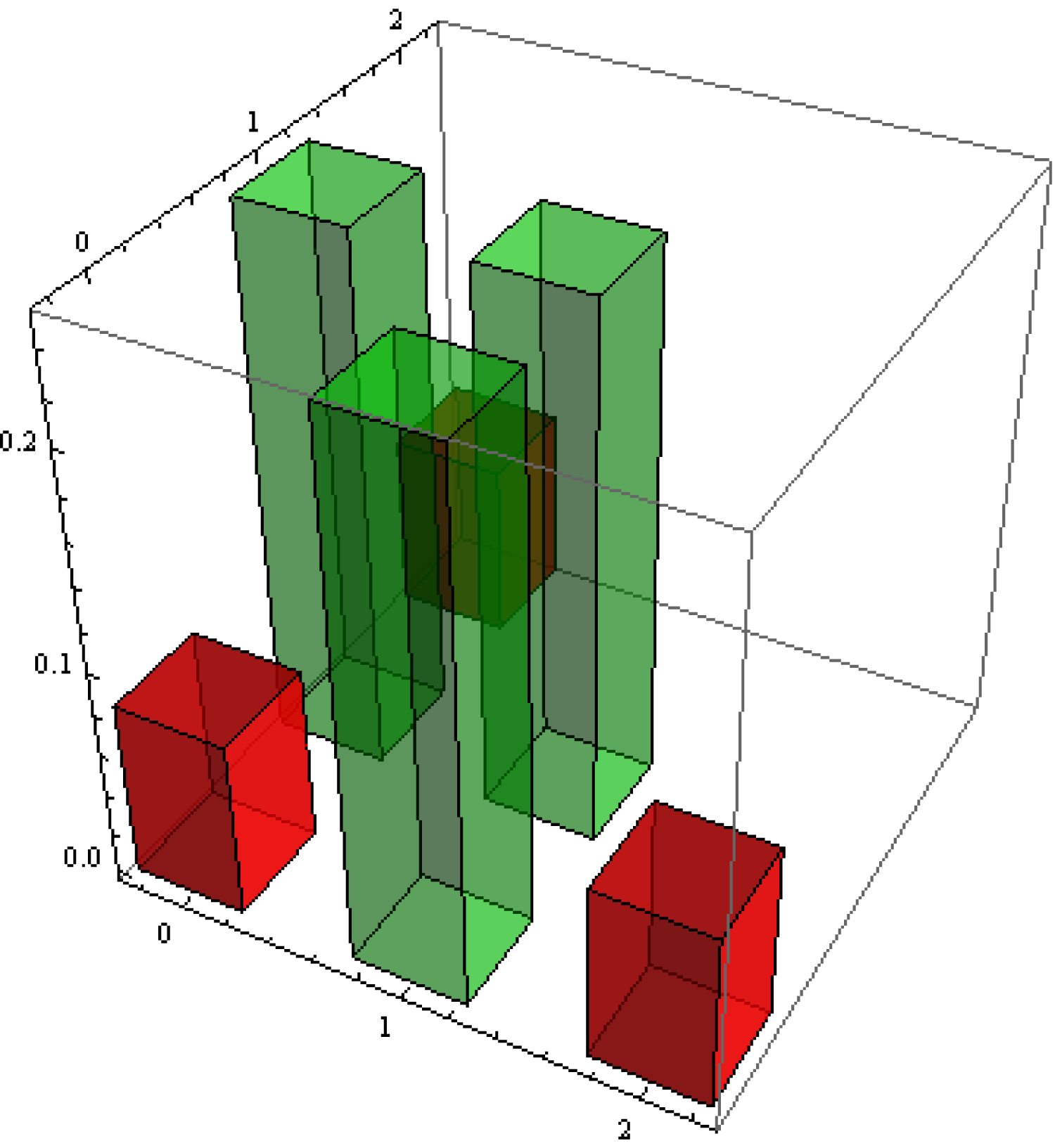}\qquad
\includegraphics[scale=0.4]{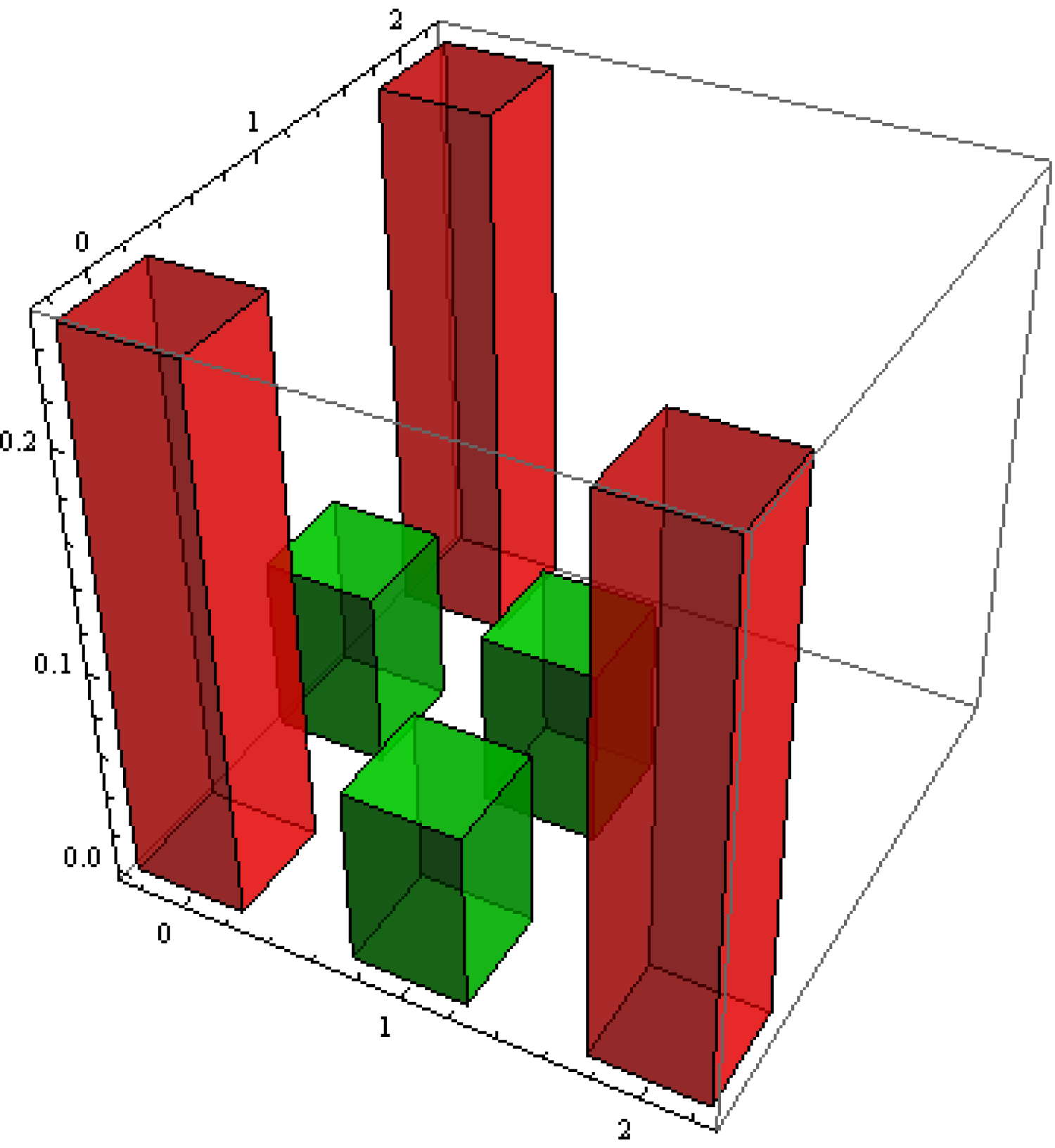}\qquad
\caption{Left: probability histogram for any one of the input states $\ket{200}, \ket{020}$ or $\ket{002}$.  Right: 
probability histogram for any one of the input states $\ket{110}, \ket{101}$ or $\ket{011}$.  
Columns of the same color have the same height.}
\label{M2histograms}
\end{center}
\end{figure}

For $n_1+n_2+n_3=2$, it is convenient to construct probability histograms for points in the first hextant.  
Using as input state any one of the states $\ket{200}\leftrightarrow (2,0)$,  $\ket{020}\leftrightarrow (0,2)$ or $\ket{002}\leftrightarrow (0,0)$, we find two and only two possible amplitudes, as 
illustrated with two different colors on the left of Fig.\ref{M2histograms}.  The corresponding histogram for any one of the input states 
 $\ket{110}\leftrightarrow (1,1) $ or $\ket{101}\leftrightarrow (1,0)$ or $\ket{011}\leftrightarrow (0,1)$ is on the right of Fig.\ref{M2histograms}.  

For a given input state not every Fourier component has the same amplitude: complementarity in the
sense of Jordan is lost - as expected since $\vert F_{(n_1n_2),(a,b)}\vert^2$ is no longer constant.  However, 
points $(n_1,n_2)$ and $(n_1',n_2')$ with equal amplitudes in the first hextant correspond to physical states $\ket{n_1n_2n_3}$ and
$\ket{n_1'n_2'n_3'}$ which differ by a permutation of their entries.  

For a generic input state, like $\ket{0,21,9}$, the probability landscape is rugged without any special features.   However, for input states of the 
type $\ket{N00}$ or $\ket{0N0}$ or $\ket{00N}$, which are mapped to the corner edges of the first hextant, we find that the probability landscape is
remarkably regular.
\begin{figure}[h!]
\begin{center}
\includegraphics[scale=0.35]{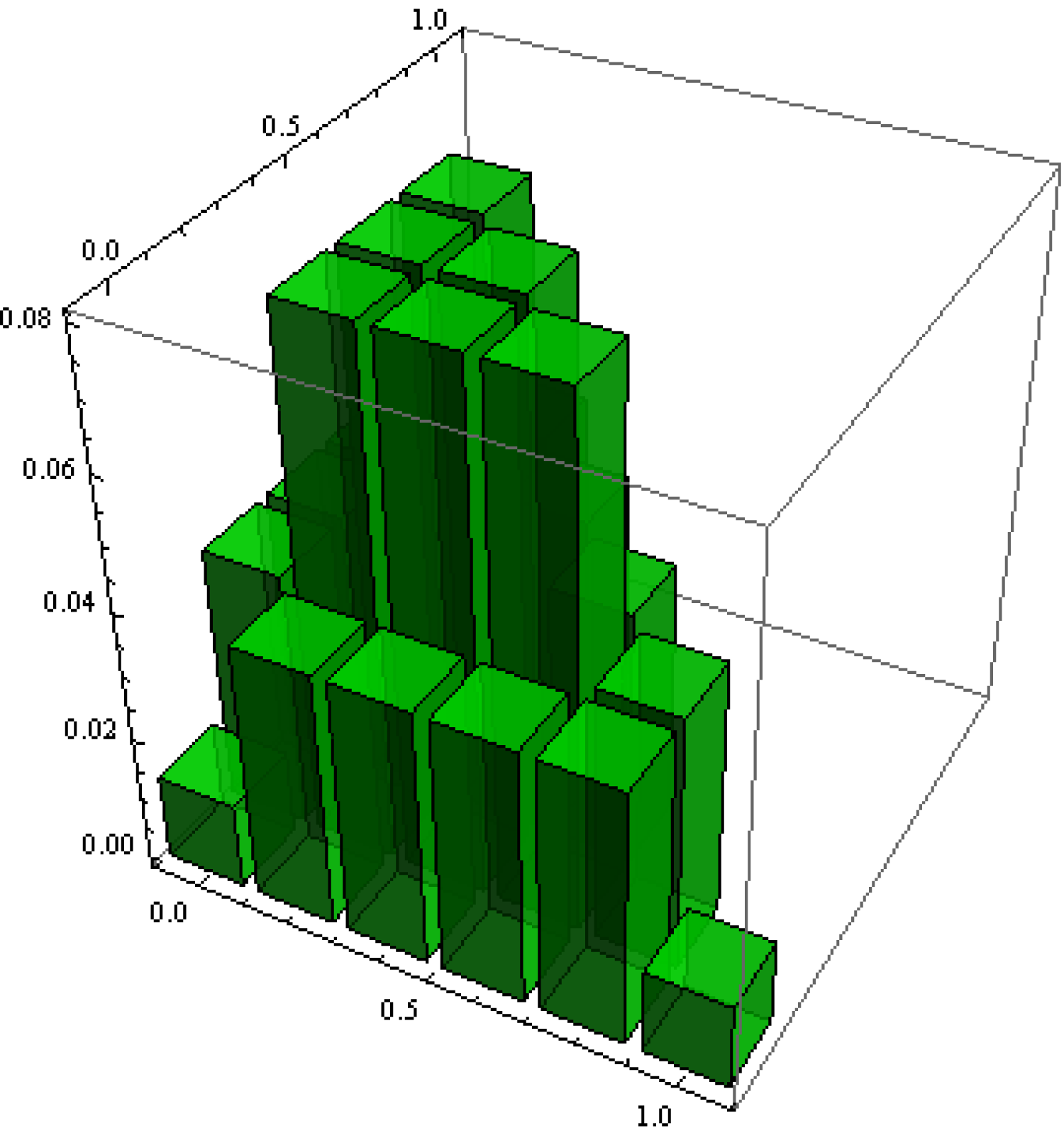}\quad\includegraphics[scale=0.35]{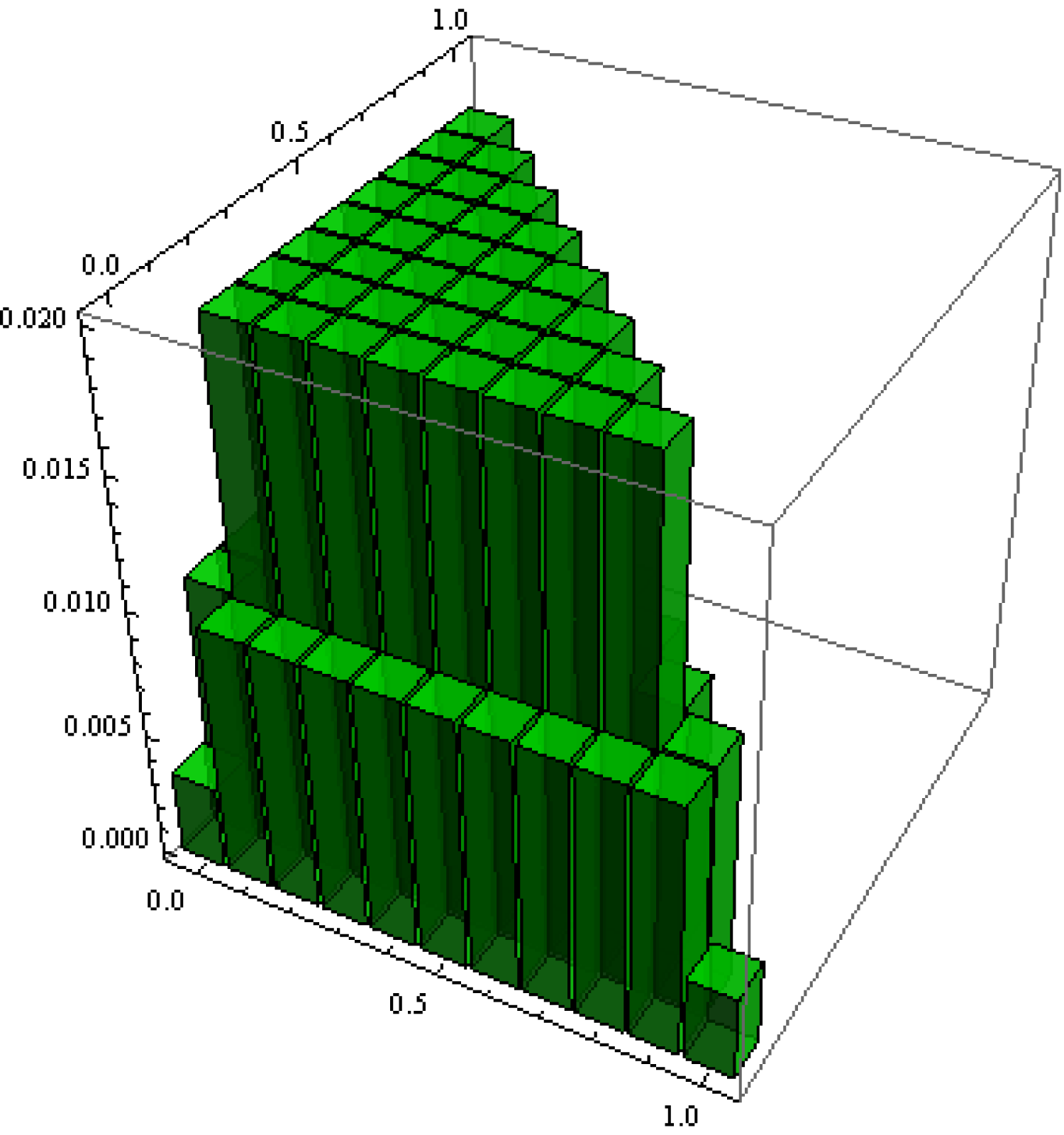}
\quad\includegraphics[scale=0.35]{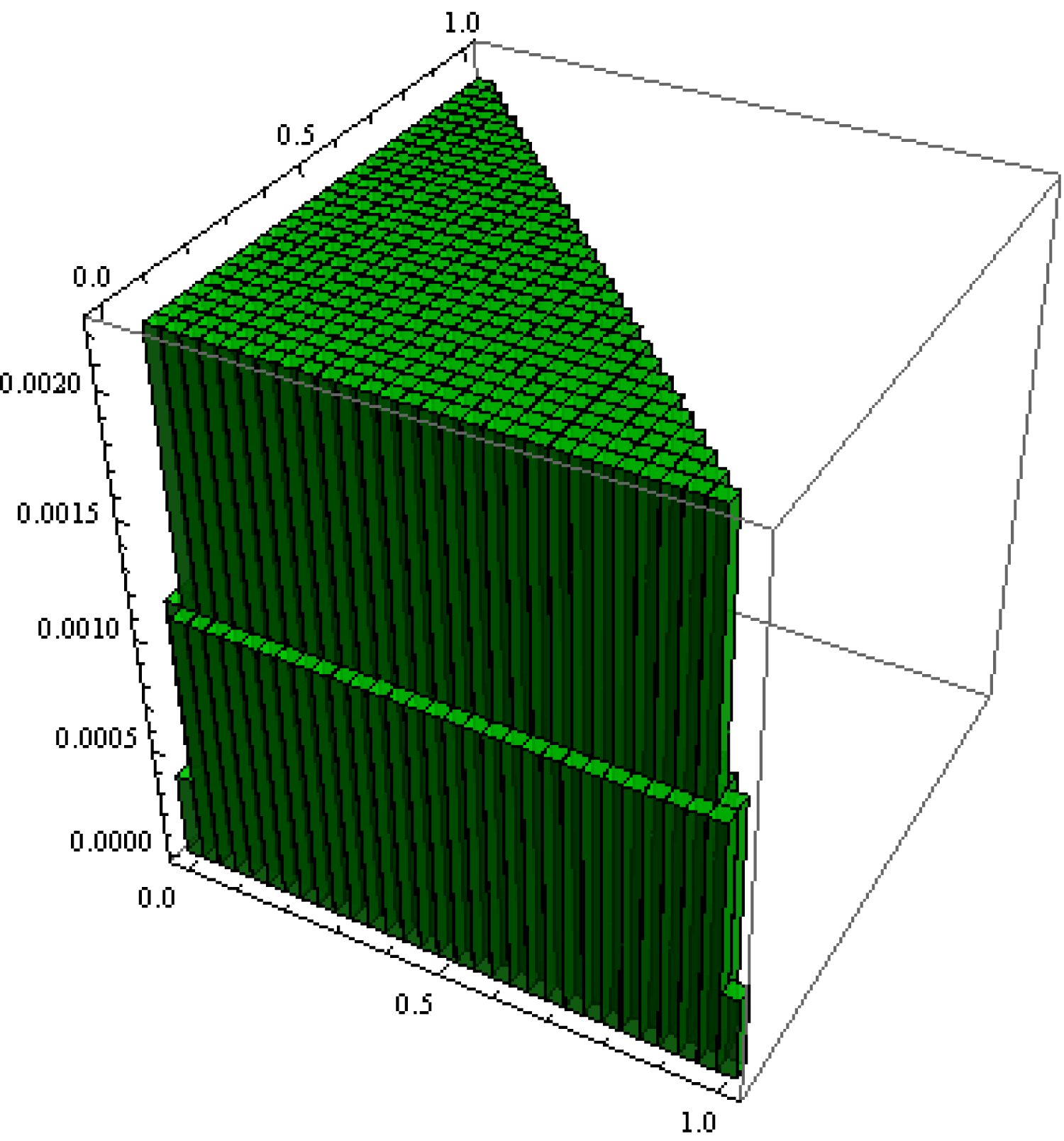}
\caption{Probability landscapes for input states $\ket{5,0,0}, \ket{10,0,0}$ and $\ket{20,0,0}$.}
\label{cornerstates}
\end{center}
\end{figure}
The probability landscape is \emph{asymptotically flat}, meaning that, in the large $N$ limit,
the phase states $F\ket{N00}$ etc are \emph{asymptotically conjugate} to the Fock states $\ket{n_1n_2n_3}$.

\subsection{$su(3)$ phase operators}

\begin{figure}[h!]
\begin{center}
\includegraphics[scale=0.4]{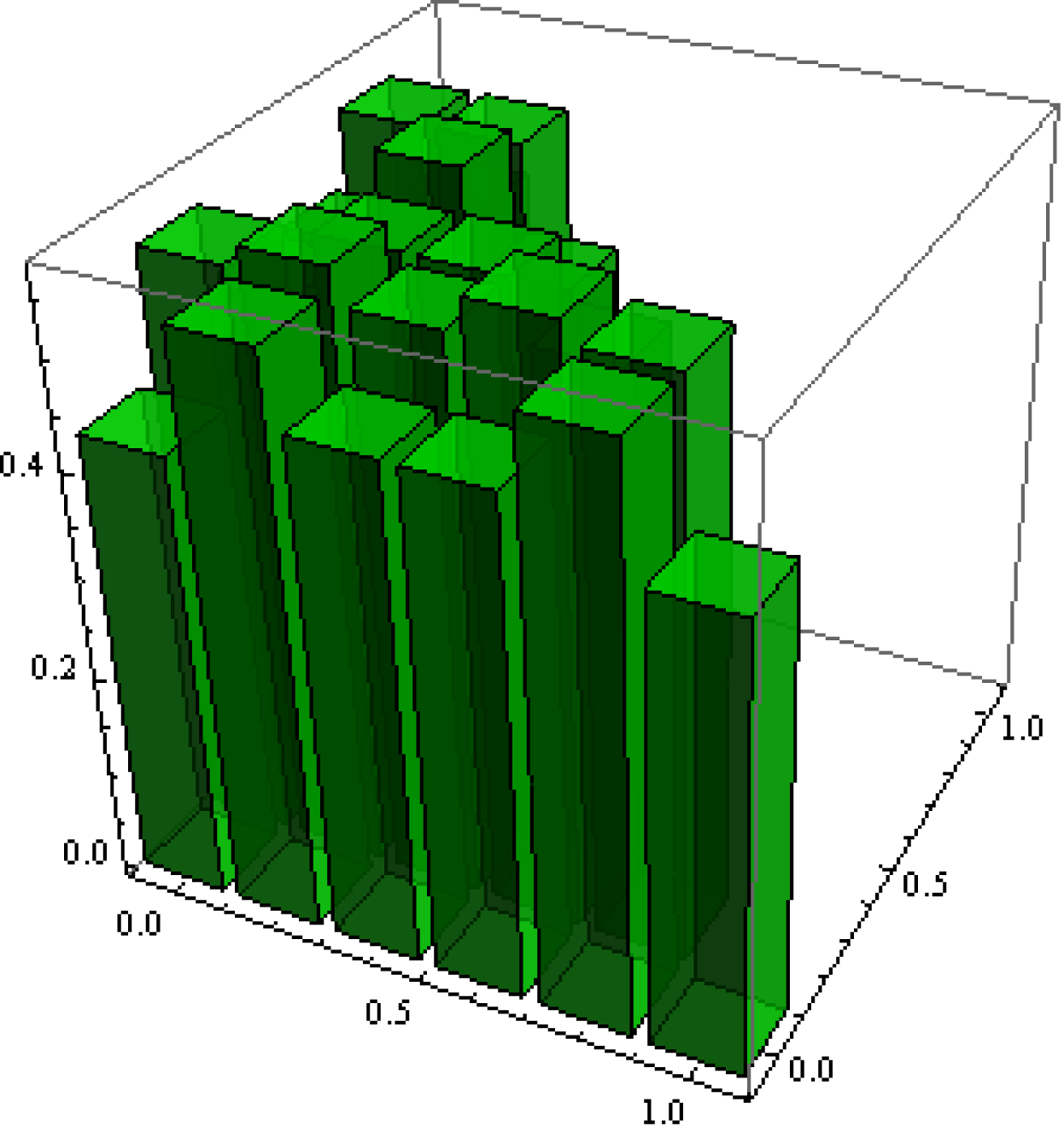}\qquad\qquad
\includegraphics[scale=0.3]{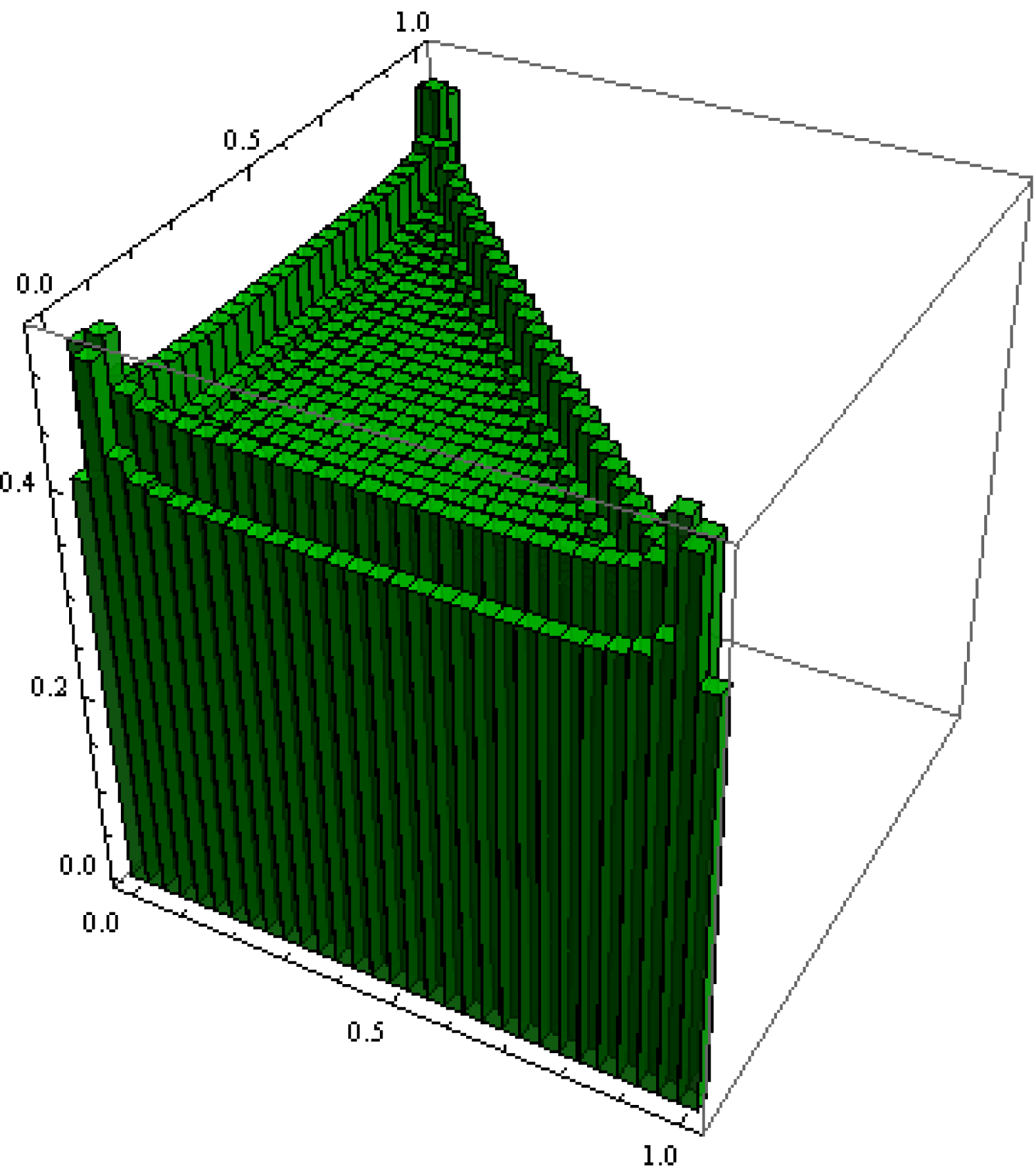}
\caption{Variance of the phase operator $\hat \eta_1$ calculated using the physical states
$\ket{n_1n_2n_3}$.  Left: $n_1+n_2+n_3=5$.  Right: $n_1+n_2+n_3=30$.}\label{variances}
\end{center}
\end{figure}

The phase operators $\hat \eta_1,\hat \eta_2$ of Eqn.(\ref{su3phaseops}) generally have ``complicated'' expressions.
In spite of this, we have found the following observation to hold.  
If we evaluate the variances $\Delta \eta_1$ and $\Delta \eta_2$ using the physical states $\ket{n_1n_2n_3}$, 
the smallest variances always occur for the states $\ket{N00}$, $\ket{0N0}$ or $\ket{00N}$.  The landscape of variances
of $\hat \eta_1$ evaluated in the physical states $\ket{n_1n_2n_3}$ is illustrated in Fig.\ref{variances} for 
$n_1+n_2+n_3=5$ and $30$.

\section{Conclusions}

The polar decomposition of operators in $SU(3)$ produces phase operators that are ambiguous and not unique:
in general, non--commuting raising operators lead to a decomposition that produces non--commuting phase operators.
Moreover, this approach produces an ``exponential phase'' rather than a phase operator.
 
We can obtain hermitian commuting ``phase--like'' operators by using of symmetry--adapted FFT.  The procedure is
mathematically systematic but not very intuitive, and we loose the connection with complementarity.   With this approach the
physical states $\ket{N00}, \ket{0N0}$ and $\ket{00N}$ stand out as having unexpected properties of asymptotic complementarity.  The
variances of the phase operators evaluated using those states is always the smallest.

An unanswered question
(not discussed in this contribution) is the difficulty in imposing correct cyclic boundary conditions on the phase operators themselves
once they are exponentiated.

This work was supported in part by NSERC of Canada and Lakehead University.

\end{document}